\documentclass[%
 reprint,
 amsmath,amssymb,
 aps, prd,
floatfix
]{revtex4-2}

\usepackage{graphicx}
\usepackage{dcolumn}
\usepackage{bm}
\usepackage{appendix}
\usepackage{xcolor}

\begin{document}


\title{Comparison of simulated neutrino emission models with data on Supernova 1987A}

\author{Jackson Olsen}
\author{Yong-Zhong Qian}
\affiliation{School of Physics and Astronomy, University of Minnesota, Minneapolis, Minnesota 55455}

\date{\today}

\begin{abstract}
We compare models of supernova (SN) neutrino emission with the Kamiokande II data on SN 1987A using the Bayesian approach.
These models are taken from simulations and are representative of current 1D SN models. We find that models with a brief
accretion phase of neutrino emission are the most favored. This result is not affected by varying the overall flux normalization
or considering neutrino oscillations. We also check the compatibility of the best-fit models with the data.
\end{abstract}

\maketitle

\section{Introduction}
\label{sec:intro}
In this paper, we compare neutrino emission models representative of current 1D supernova (SN) simulations
with the data on SN 1987A. Approximately twenty neutrino events were observed by the Kamiokande II (KII) 
\cite{PhysRevLett.58.1490, PhysRevD.38.448}, Irvine-Michigan-Brookhaven (IMB) \cite{PhysRevLett.58.1494}, 
and Baksan \cite{ALEXEYEV1988209} detectors. These events have been extensively studied to understand
SN neutrino emission and neutrino properties (see e.g., \cite{arnett1989} for a review of earlier works 
and e.g., \cite{Loredo:2001rx} and \cite{Costantini_2007} for detailed methodical analyses). The common practice was to use 
parametric models of SN neutrino emission and extract several parameters from 
the relatively sparse data for comparison with the results from SN neutrino transport calculations.
These calculations have been greatly refined over the last several decades, during which there have been major 
advances in SN modeling (see e.g., \cite{janka2012,Mirizzi:2015eza} for reviews). Rather than approximating
these detailed SN neutrino emission models with some simplified parameterization, it is interesting to 
compare them directly with the SN 1987A data. We follow \cite{Loredo:2001rx} and use the Bayesian approach 
for such comparisons in this paper. This approach provides a straightforward way to rank the models in
light of the data. We also check the compatibility of the best-fit models with the data.

We take three baseline models from the Garching group \cite{Garching}: z9.6-SFHo, s20-SFHo, and s27-LS220,
which will be referred to as models A, B, and C, respectively. Models A and C were described in detail
in \cite{Mirizzi:2015eza}. These three models are representative of current 1D SN simulations and cover
a range of neutrino emission for comparison with the SN 1987A data. We also vary the overall flux 
normalization to obtain models A$'$, B$'$, and C$'$ as counterparts of models A, B, and C and consider 
three cases of neutrino oscillations for each model. In total, we compare eighteen models with the
SN 1987A data from the KII detector. We focus on the eleven neutrino events observed in this detector because the IMB detector had issues of 
failing photomultiplier tubes \cite{PhysRevLett.58.1494}, and therefore, is harder to characterize while 
the Baksan detector reported significantly fewer events. We find that models with a brief accretion 
phase of neutrino emission are the most favored. This result is not affected 
by varying the overall flux normalization or considering neutrino oscillations.

This paper is organized as follows. We describe our adopted SN neutrino emission models in Sec.~\ref{sec:model}
and our Bayesian approach to compare them with the data in Sec.~\ref{sec:method}. We present the results 
in Sec.~\ref{sec:result} and check the compatibility of the best-fit models with the data in Sec.~\ref{sec:pvalue}. 
We summarize our results and give conclusions in Sec.~\ref{sec:discuss}.

\section{Neutrino Emission Models}
\label{sec:model}
In general, the neutrino emission relevant for comparison with the SN 1987A data
consists of an accretion phase followed by a much longer cooling phase. During the accretion phase, material
is still falling through the standing SN shock onto the proto-neutron star (PNS) and the release of the 
gravitational binding energy of this material gives rise to enhanced emission of $\nu_e$ and $\bar\nu_e$
over $\nu_x$ and $\bar\nu_x$ $(x=\mu,\ \tau)$ through $e^-+p\to n+\nu_e$ and $e^++n\to p+\bar\nu_e$. 
The duration of this phase depends on the details of the explosion. It is brief for a light progenitor 
such as the one of $9.6\,M_\odot$ in model A, where the density of the infalling material rapidly decreases 
with radius and consequently the shock experiences only a little hindrance in moving out. In contrast, 
the shock has a much harder time overcoming the infalling material in more massive progenitors such as 
those of 20 and $27\,M_\odot$ in models B and C, respectively, and therefore, there are extended accretion 
phases of neutrino emission in these models. For all cases with a stable PNS, most of the gravitational 
binding energy is carried away during the cooling phase, when all neutrino species have approximately 
the same luminosity. The gravitational binding energy of the PNS depends on both the progenitor and the nuclear
equation of state (EoS). The SFHo EoS from \cite{SFHo} was used for models A and B, and the LS220 EoS from
\cite{LS220} with a compression modulus of 220 MeV was used for model C.

In the absence of flavor oscillations, the energy-differential number flux of a neutrino species $\nu_\beta$ 
at a distance $d$ from its SN source is
\begin{equation}
    F_{\nu_\beta}(E_\nu,t) = \frac{L_{\nu_\beta}}{4 \pi d^2\langle E_{\nu_\beta}\rangle}f_{\nu_\beta}(E_\nu,t),
\end{equation} 
where $L_{\nu_\beta}$ and $\langle E_{\nu_\beta}\rangle$ are the $\nu_\beta$ luminosity and average energy,
respectively, and $f_{\nu_\beta}(E_\nu,t)$ is the normalized $\nu_\beta$ energy spectrum. This spectrum can be 
fitted to the form (e.g., \cite{Tamborra2012})
\begin{equation}
f_{\nu_\beta}(E_\nu,t)=\frac{T^{-1}_{\nu_\beta}(t)}{\Gamma(1 + \alpha_{\nu_\beta}(t))}
\left(\frac{E_\nu}{T_{\nu_\beta}(t)}\right)^{\alpha_{\nu_\beta}(t)}
e^{-E_\nu/T_{\nu_\beta}(t)},
\end{equation}
where $\Gamma(1 + \alpha_{\nu_\beta})$ refers to the Gamma function, and $\alpha_{\nu_\beta}$ and $T_{\nu_\beta}$ 
are related to the first and second moments of $E_\nu$ for the $\nu_\beta$ spectrum from SN simulations by
\begin{equation}
\alpha_{\nu_\beta}=\frac{2 \langle E_{\nu_\beta} \rangle^2 -  \langle E_{\nu_\beta}^2 \rangle}
{\langle E_{\nu_\beta}^2 \rangle - \langle E_{\nu_\beta} \rangle^2}
\end{equation}
and
\begin{equation}
T_{\nu_\beta}=\frac{\langle E_{\nu_\beta}\rangle}{1 + \alpha_{\nu_\beta}}.
\end{equation}
Note that $L_{\nu_\beta}$, $\langle E_{\nu_\beta}\rangle$, $\langle E_{\nu_\beta}^2 \rangle$, $\alpha_{\nu_\beta}$,
and $T_{\nu_\beta}$ are functions of time, although we usually suppress their time dependence for convenience.

\begin{figure}
\includegraphics[width=\columnwidth]{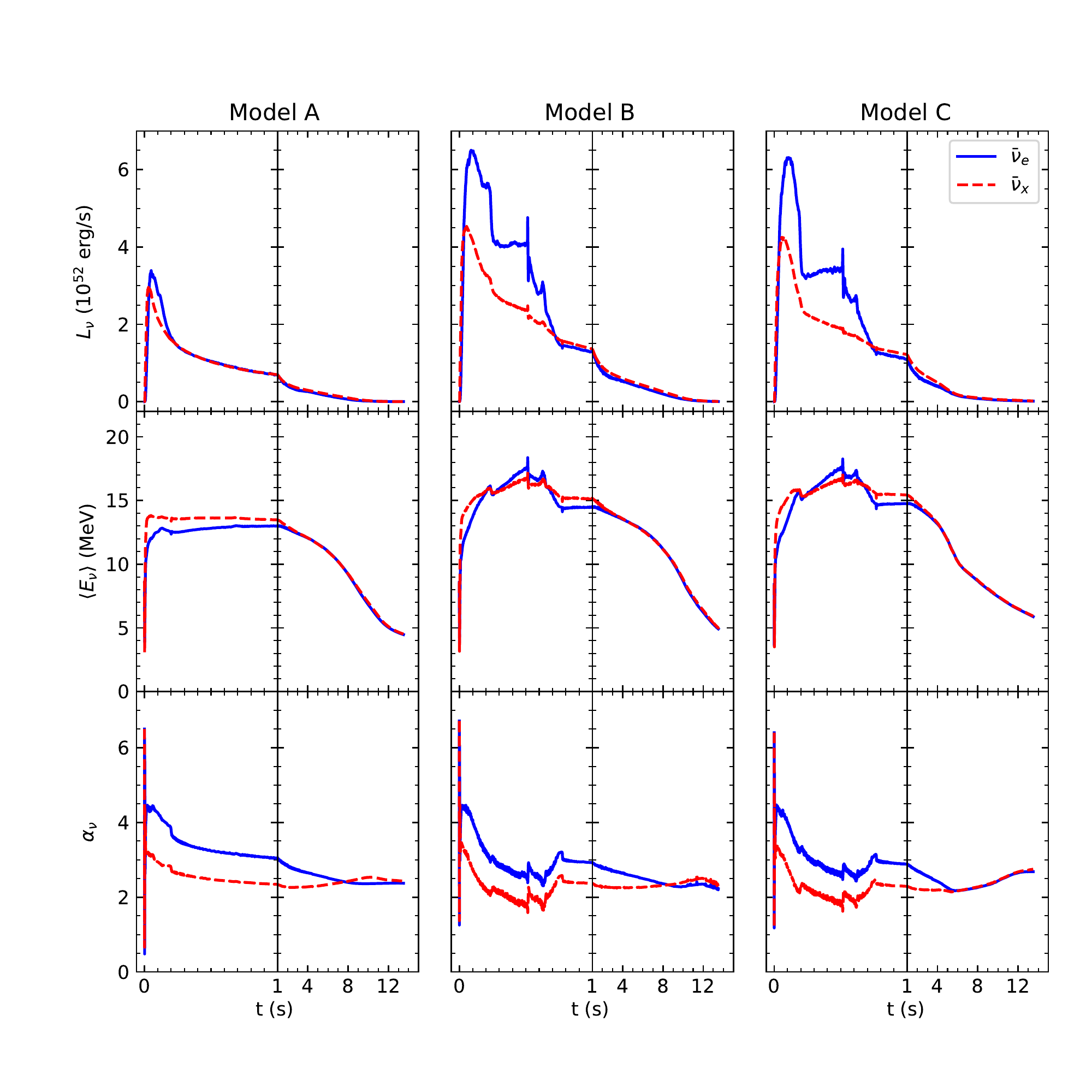}
\caption{The luminosity $L_{\nu}$, average energy $\langle E_{\nu}\rangle$, and spectral parameter 
$\alpha_{\nu}$ are shown as functions of time for $\bar\nu_e$ and $\bar\nu_x$ for the baseline models A, B, and C.
Note the change of the time scale at $t=1$~s.}
\label{fig:emission}
\end{figure}

Because the SN neutrino events in the KII detector were induced predominantly by $\bar\nu_e$, we show in
Fig.~\ref{fig:emission} the time evolution of $L_{\bar\nu_e}$, $\langle E_{\bar\nu_e}\rangle$, and
$\alpha_{\bar\nu_e}$ for models A, B, and C. We also show the same information on $\bar\nu_x$ for 
consideration of neutrino oscillations. A duration of 13.5~s is chosen for comparison with the SN 1987A data.
It can be seen from Fig.~\ref{fig:emission} that the turn-on of neutrino emission is extremely rapid for all models.
We focus on the emission features subsequent to the turn-on below. 

As mentioned above, models B and C have much more extended accretion phases (with significant excess of 
$L_{\bar\nu_e}$ over $L_{\bar\nu_x}$) than model A. The former two models also have higher $L_{\bar\nu_e}$
and $L_{\bar\nu_x}$ than the latter. The total energy $\cal{E}_\nu$
emitted in all neutrino species for each model is given in Table~\ref{tab:models}.
For all the models, $\langle E_{\bar\nu_e}\rangle$ and $\langle E_{\bar\nu_x}\rangle$ have some differences
for the first $\sim 1$~s but are nearly the same later on. The difference between $\alpha_{\bar\nu_e}$
and $\alpha_{\bar\nu_x}$ persists at least up to $t\sim 5$~s but also diminishes at late times.
In addition, the $\langle E_{\bar\nu_e}\rangle$ and $\langle E_{\bar\nu_x}\rangle$ for model B are significantly 
larger than those for model A. Because these two models employ the same EoS, their $\langle E_{\bar\nu_e}\rangle$ 
and $\langle E_{\bar\nu_x}\rangle$ have nearly constant differences for most of the evolution shown in 
Fig.~\ref{fig:emission}. For model C with a different EoS, its $\langle E_{\bar\nu_e}\rangle$ and 
$\langle E_{\bar\nu_x}\rangle$ are close to those for model B up to $t\sim 4$~s. Subsequently, compared to
model B, the $\langle E_{\bar\nu_e}\rangle$ and $\langle E_{\bar\nu_x}\rangle$ for model C decrease more 
rapidly for a brief period and then decrease more slowly.

\begin{table}[tbh]
\caption{Baseline models.}
    \centering
    \begin{tabular}{cccc}
        \hline
        \hline
        Model & progenitor mass ($M_\odot$) & EoS & $\cal{E}_\nu$ ($10^{53}$ ergs)\\
        \hline
        A & 9.6 & SFHo & 1.99 \\
        B & 20 & SFHo & 4.28 \\
        C & 27 & LS220 & 3.30 \\
        \hline
        \hline
    \end{tabular}
    \label{tab:models}
\end{table}

Models A, B, and C were calculated for three different progenitors and employ two different forms of EoS; we consider that they cover a range of neutrino emission relevant for the case of SN 1987A.
To study models with additional freedom, we also consider models A$'$, B$'$, and C$'$ with scaled fluxes
\begin{equation}
    F_{\nu_\beta}'(E_\nu,t) = \frac{KL_{\nu_\beta}}{4 \pi d_0^2\langle E_{\nu_\beta}\rangle}f_{\nu_\beta}(E_\nu,t),
\end{equation}
where $K$ is a scale factor and $d_0=51.4$~kpc is the
central value of the measured distance to SN~1987A \cite{1991ApJ...380L..23P,Panagia:2003rt}. 
The scale factor $K$ is the same for all neutrino species and allows the total emitted neutrino energy $\cal{E}_\nu$ to vary.

The $\bar\nu_e$ flux at the detector is modified by flavor oscillations. We consider three cases to sample
these effects. Specifically, for models A, B, and C, the detected $\bar\nu_e$ flux is taken to be
\begin{equation}
    F_{\rm det}(E_\nu,t) = f F_{\bar\nu_e}(E_\nu,t) + (1-f) F_{\bar\nu_x}(E_\nu,t),
\end{equation}
where the constant $f$ specifies the degree of mixing between $\bar\nu_e$ and $\bar\nu_x$.
The reference case of no oscillations (NO) corresponds to $f=1$. The other two cases correspond to
$f = 0.681$ or 0.022 for just Mikheyev-Smirnov-Wolfenstein flavor transformation with the normal (NH) or 
inverted (IH) neutrino mass hierarchy, respectively \cite{PhysRevD.62.033007, Gonzalez-Garcia, Zyla:2020zbs}. 
The same three cases are also considered for models A$'$, B$'$, and C$'$. So altogether we compare eighteen
models of SN neutrino emission with the SN 1987A data. For convenience, we add (NO), (NH), or (IH) to label
a model when neutrino oscillations are of concern. For example, model A (NO) denotes model A with no oscillations,
and  model A (NH) denotes model A including oscillations with the NH.

\section{Bayesian Approach to Compare Models with Data}
\label{sec:method}
The KII detector consisted of 2.14 kton of water in its fiducial volume and observed the neutrinos from SN 1987A 
predominantly through the Cherenkov radiation of the $e^+$ produced by inverse beta decay (IBD)
$\bar\nu_e+p\to n+e^+$. Below we assume that all the neutrino events were due to IBD.
The expected energy-differential rate of events including both the signal and the background is
\begin{equation}
\begin{split}
    \frac{d^2N}{dtdE}(E,t) = B(E) + N_p\int F_{\rm det}(E_\nu,t)\sigma_{\rm IBD}(E_\nu)\\
    \times\frac{\epsilon(E_e)}{\sigma_E\sqrt{2 \pi}}
    \exp\left[-\frac{(E-E_e)^2}{2 \sigma_E^2}\right]dE_\nu,
\end{split}
\label{eq:eventrate}
\end{equation}
where $B(E)$ is the background rate at energy $E$ taken from \cite{Loredo:2001rx} but treated as in \cite{Costantini_2007}, 
$N_p$ is the total number of free protons within the fiducial volume, $\sigma_{\rm IBD}(E_\nu)$ is the IBD cross section, 
$E_e=E_\nu-\Delta$ is the energy of the $e^+$ from the IBD reaction, $\Delta=1.293$~MeV is the neutron-proton mass difference, and 
$\epsilon(E_e)$ is the detection efficiency taken from \cite{Loredo:2001rx}. Due to smearing, an $e^+$ of energy $E_e$ may be 
detected at energy $E$, the probability of which is approximated by a Gaussian distribution with 
a standard deviation $\sigma_E = \sqrt{(0.75 ~\text{MeV}) E_e}$
\cite{PhysRevD.54.1194, 55af5131dc954e7592bb1d948c16becd} in Eq.~(\ref{eq:eventrate}).

Table~\ref{tab:events} lists the detection time $t_{\rm det}$ and the detected energy $E$ 
for each of the eleven KII events with $E\geq 7.5$~MeV for SN 1987A.
The energy cutoff is chosen so that the models and the data are
compared only for $t\leq 13.5$~s (there were four events below the cutoff during $t_{\rm det}=17.641$--23.814~s \cite{Loredo:2001rx}). 
The first event is defined by $t_{\rm det}=0$. Due to the random nature of detection, there is a time offset $t_{\rm off}$ 
between the first event and the model time $t=0$ 
(the time of travel from SN 1987A to the detector is the same for all the events, and therefore, can be ignored). 
So $t=t_{\rm det}+t_{\rm off}$. 

\begin{table}[tbh]
\caption{KII data for SN 1987A.}
    \centering
    \begin{tabular}{cc}
    \hline
    \hline
    $t_{\rm det}$ (s) & $E$ (MeV) \\
    \hline
    0 & 20.0 \\ 
    0.107 & 13.5 \\
    0.303 & 7.5 \\
    0.324 & 9.2 \\
    0.507 & 12.8 \\
    1.541 & 35.4 \\
    1.728 & 21.0 \\
    1.915 & 19.8 \\
    9.219 & 8.6 \\
    10.433 & 13.0 \\
    12.439 & 8.9 \\
    \hline
    \hline
    \end{tabular}
    \label{tab:events}
\end{table}

For a specific model $M_i$ with a set of parameters $\boldsymbol{\theta}$, the probability 
of an event being detected with energy $E$ at time $t_{\rm det}=t-t_{\rm off}$ is
\begin{equation}
    p(E,t|\boldsymbol{\theta},M_i)=\frac{1}{\langle N\rangle}\frac{d^2N}{dtdE},
    \label{eq:pet}
\end{equation}
where $d^2N/dtdE$ refers to the expected energy-differential rate of events for model $M_i$, and
\begin{equation}
    \langle N\rangle=\int_0^{\rm 13.5~s}dt\int_{\rm 7.5~MeV}^\infty dE \frac{d^2N}{dtdE}
\end{equation}
is the expected total number of events. The likelihood of detecting a particular configuration of $N=11$ events is
\begin{equation}
    p(D|\boldsymbol{\theta},M_i)=\frac{\langle N\rangle^Ne^{-\langle N\rangle}}{N!}
    \prod_{j=1}^{N}p(E_j,t_j|\boldsymbol{\theta},M_i),
    \label{eq:likelihood}
\end{equation}
where $D$ represents the set of detection data, and $E_j$ and $t_j$ correspond to $E$ and $t$ for the $j$th event.
The likelihood in Eq.~(\ref{eq:likelihood}) follows from the extended maximum likelihood method \cite{BARLOW1990496}.

For all the models, the time offset $t_{\rm off}$ is a parameter. The other parameter is the distance $d$ to SN 1987A
for models A, B, and C or the scale factor $K$ for models A$'$, B$'$, and C$'$. Our simple treatment of neutrino 
oscillations does not introduce any new parameters. In the Bayesian approach, the posterior probability for
the parameters of model $M_i$ is
\begin{equation}
    p(\boldsymbol{\theta}|D,M_i)=\frac{p(D|\boldsymbol{\theta},M_i)p(\boldsymbol{\theta}|M_i)}{p(D|M_i)},
    \label{eq:post}
\end{equation}
where $p(\boldsymbol{\theta}|M_i)$ is the prior probability for the parameters and
\begin{equation}
    p(D|M_i)=\int p(D|\boldsymbol{\theta},M_i)p(\boldsymbol{\theta}|M_i)d\boldsymbol{\theta}.
    \label{eq:evid}
\end{equation}
The prior probability $p(\boldsymbol{\theta}|M_i)$ is 
$p(d|M_i)p(t_{\rm off}|M_i)$ or $p(K|M_i)p(t_{\rm off}|M_i)$. We take $p(d|M_i)$ to be a Gaussian distribution
with a mean of 51.4~kpc and a standard deviation of 1.2~kpc based on the measurement in \cite{Panagia:2003rt}, 
and adopt uniform distributions for $p(t_{\rm off}|M_i)$ and $p(K|M_i)$.

The Bayesian approach can also be applied to obtain the (discrete) posterior probability for model $M_i$ as
\begin{equation}
    p(M_i|D)=\frac{p(D|M_i)p(M_i)}{p(D)},
    \label{eq:evidm}
\end{equation}
where $p(M_i)$ is the prior probability for model $M_i$ and
\begin{equation}
    p(D)=\sum_ip(D|M_i)p(M_i).
    \label{eq:evidence}
\end{equation}
Including three cases of neutrino oscillations, we take $p(M_i)=1/9$ and consider the set of models A, B, and C 
separately from the set of their counterparts A$'$, B$'$, and C$'$. For comparing models $M_i$ and $M_j$ in a set, 
it is convenient to introduce the Bayes factor
\begin{equation}
    B_{ij}=p(D|M_i)/p(D|M_j),
\end{equation}
which is numerically the same as $p(M_i|D)/p(M_j|D)$ in our case.
For $B_{ij}=1$--3, 3--20, 20--150, and $>150$, the evidence in favor of model $M_i$ over model $M_j$ is 
hardly noticeable, positive, strong, and very strong, respectively \cite{Loredo:2001rx}.

\section{Results from Bayesian Approach}
\label{sec:result}
\begin{figure}
\includegraphics[width=\columnwidth]{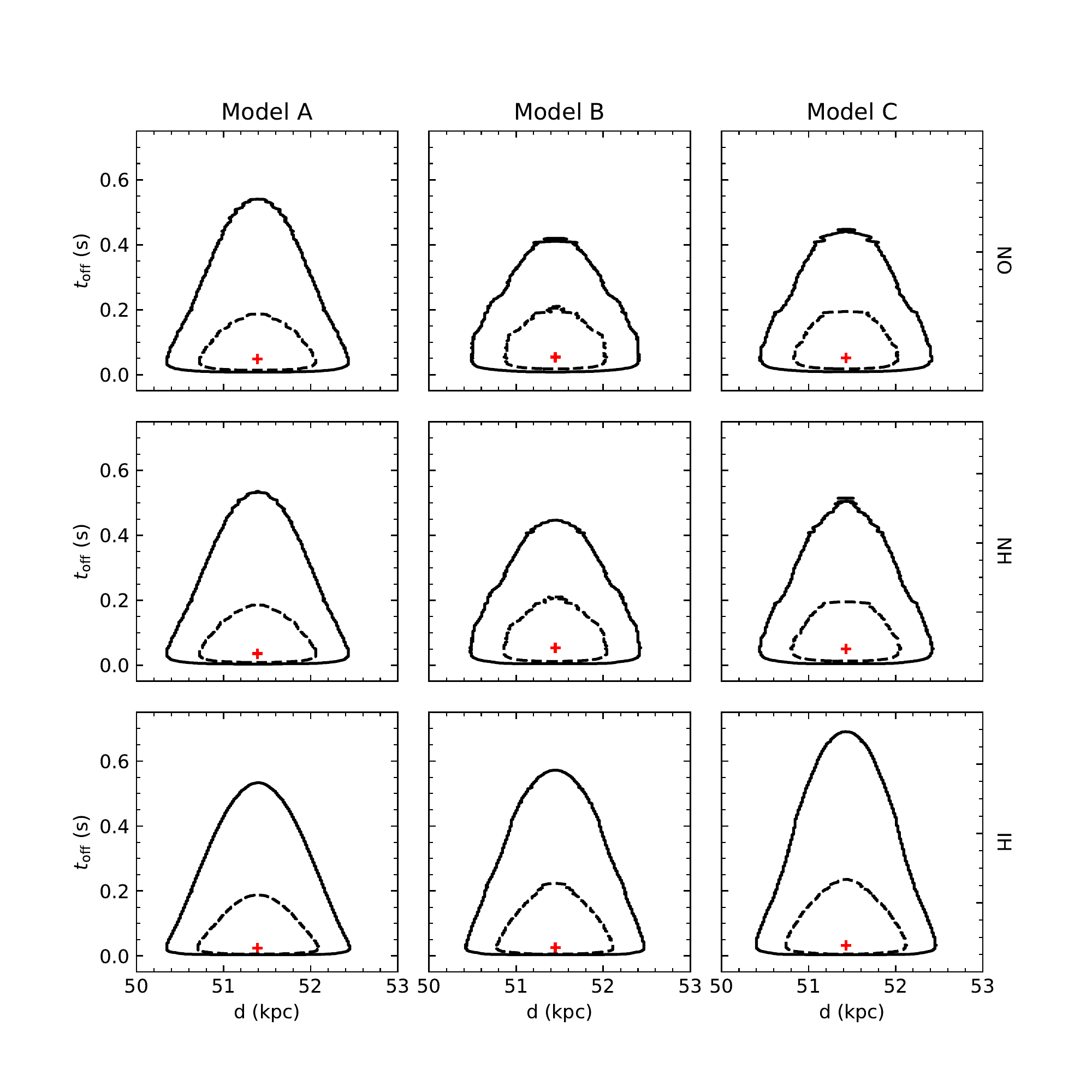}
\caption{The best-fit values (plus) and the 68\% (dashed curve) and 95\% (solid curve) credible
regions for $d$ and $t_{\rm off}$ are shown for models A, B, and C including three cases of neutrino oscillations.}
\label{fig:result-dt}
\end{figure}

\begin{figure}
\includegraphics[width=\columnwidth]{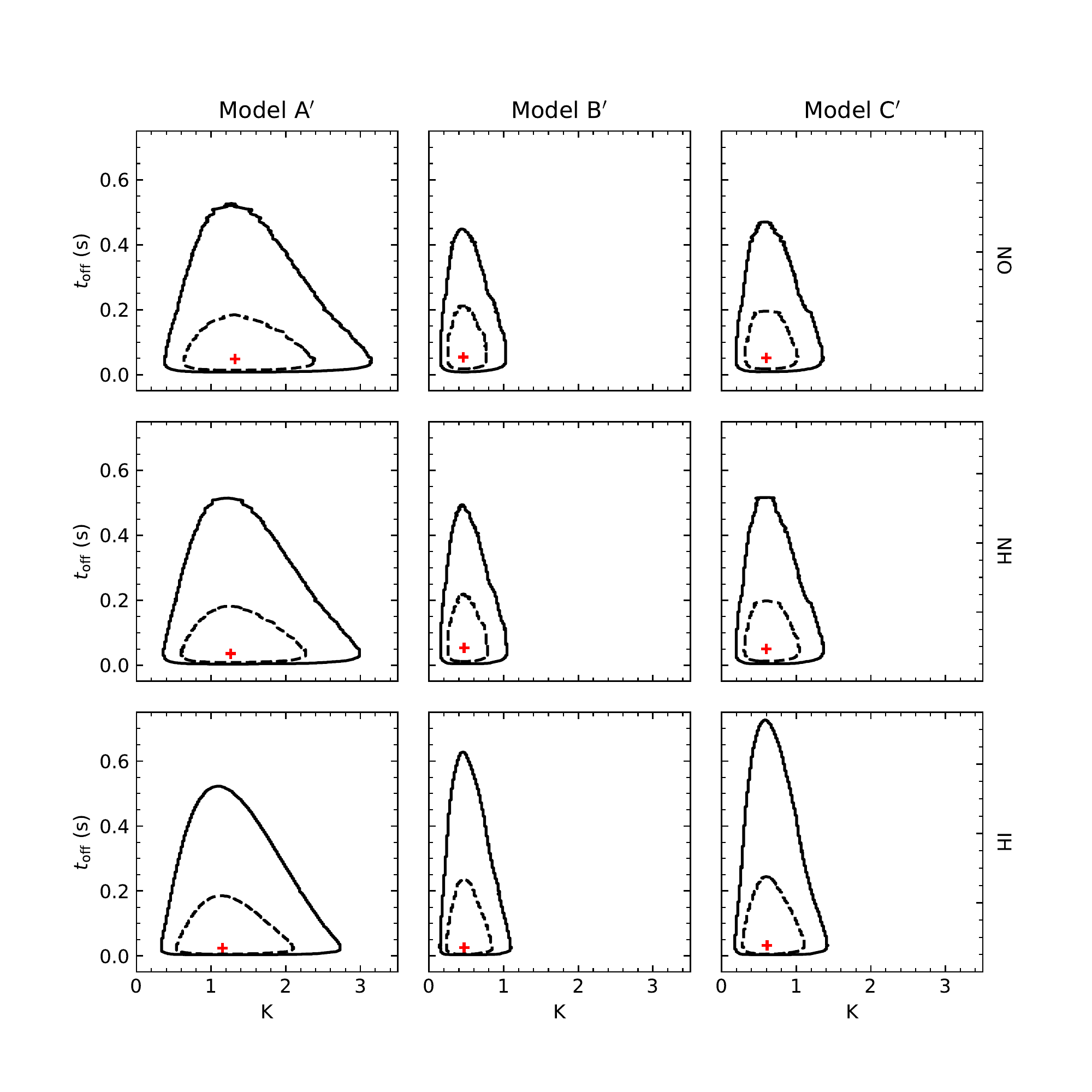}
\caption{The best-fit values (plus) and the 68\% (dashed curve) and 95\% (solid curve) credible
regions for $K$ and $t_{\rm off}$ are shown for models A$'$, B$'$, and C$'$ including three cases of neutrino oscillations.}
\label{fig:result-kt}
\end{figure}

Using the posterior probability in Eq.~(\ref{eq:post}), we calculate the best-fit values and 
the 68\% and 95\% credible regions for the parameters. These results are shown in 
Table~\ref{tab:result} (\ref{tab:resultp}) and Fig.~\ref{fig:result-dt} (\ref{fig:result-kt})
for models A, B, and C (A$'$, B$'$, and C$'$) including three cases of neutrino oscillations. 
Tables~\ref{tab:result} and \ref{tab:resultp} also give the expected total number of events 
$\langle N\rangle$ corresponding to the best-fit parameters and the posterior probability 
$p(M_i|D)$ [see Eq.~(\ref{eq:evidm})] for each model. 

\begin{table}[tbh]
\caption{Best-fit values of $d$ and $t_{\rm off}$ and the corresponding $\langle N \rangle$
for models A, B, and C including three cases of neutrino oscillations, along with the posterior 
probability for each model.}
\centering
\begin{tabular}{lccccc}
\hline \hline
    Model & $d$ (kpc) & $t_{\rm off}$ (s) & $\langle N \rangle$ & $p(M_i|D)$ \\
    \hline
    A (NO)& 51.39& 0.048 & 6.81 & 0.2807 \\
    A (NH) & 51.39& 0.036 & 7.17 & 0.2684 \\
    A (IH) & 51.39& 0.024 & 7.92 & 0.2037 \\
    B (NO)& 51.45& 0.054 & 19.5 & 0.0058 \\
    B (NH) & 51.45& 0.054 & 19.4 & 0.0060 \\
    B (IH) & 51.45& 0.026 & 19.3 & 0.0043 \\
    C (NO)& 51.43& 0.051 & 15.1 & 0.0913 \\
    C (NH) & 51.43& 0.051 & 15.1 & 0.0875 \\
    C (IH) & 51.43& 0.033 & 15.0 & 0.0523 \\
    \hline \hline
\end{tabular}
\label{tab:result}
\end{table}

\begin{table}[tbh]
\caption{Best-fit values of $K$ and $t_{\rm off}$ and the corresponding $\langle N \rangle$
and $p$-value for models A$'$, B$'$, and C$'$ including three cases of neutrino oscillations, 
along with the posterior probability for each model.}
\centering
\begin{tabular}{lccccc}
\hline \hline
    Model & $K$ & $t_{\rm off}$ (s) & $\langle N \rangle$ & $p$-value & $p(M_i|D)$ \\
    \hline
    A$'$ (NO)& 1.32 & 0.048 & 8.90 & 0.16 & 0.2638 \\
    A$'$ (NH)& 1.26 & 0.036 & 8.96 & 0.19 & 0.2223 \\
    A$'$ (IH)& 1.15 & 0.024 & 9.06 & 0.25 & 0.1359 \\
    B$'$ (NO)& 0.46 & 0.054 & 9.14 & 0.19 & 0.0400 \\
    B$'$ (NH)& 0.47 & 0.054 & 9.29 & 0.25 & 0.0385 \\
    B$'$ (IH)& 0.47 & 0.026 & 9.21 & 0.39 & 0.0245 \\
    C$'$ (NO)& 0.60 & 0.051 & 9.18 & 0.12 & 0.1108 \\
    C$'$ (NH)& 0.60 & 0.051 & 9.16 & 0.17 & 0.1038\\
    C$'$ (IH)& 0.61 & 0.033 & 9.27 & 0.25 & 0.0603\\
    \hline \hline
\end{tabular}
\label{tab:resultp}
\end{table}

For models A, B, and C, because we have used the prior probability $p(d|M_i)$ based on the distance 
measurement, the best-fit values of $d$ are essentially the same as the measured central value of 
51.4~kpc regardless of neutrino oscillations. In addition, from the rapid turn-on of the neutrino 
luminosities (see Fig.~\ref{fig:emission}), we expect little offset between the start of neutrino 
emission and the detection of the first $\bar\nu_e$ event, which is confirmed by the small
best-fit values of $t_{\rm off}=0.024$--0.054~s for these models. On the other hand,
the 95\% credible regions enclose variations of $\sim\pm 1$~kpc in $d$ ($\sim\pm 1\sigma$ 
measurement error) and values of $t_{\rm off}$ as large as 0.42~s for model B (NO)
to 0.7~s for model C (IH).

Models A$'$, B$'$, and C$'$ differ from models A, B, and C, respectively, only in the overall
normalization of the neutrino fluxes. Each model of the former set has the same best-fit 
value of $t_{\rm off}$ as its counterpart in the latter set because they have the same time 
profile of neutrino emission. On the other hand, although both the scale factor $K$ and the
distance $d$ change the overall flux normalization, the neutrino fluxes corresponding to the
best-fit values of $K$ differ significantly from those corresponding to the best-fit values of
$d$ (equivalent to $K\approx 1$). These differences are caused by the different prior 
probabilities $p(K|M_i)$ and $p(d|M_i)$, with the former being much less restrictive than the latter.
Likewise, the $K$ values enclosed by the 95\% credible regions for models A$'$, B$'$, and C$'$
correspond to much larger variations of the neutrino fluxes than the $d$ 
values for models A, B, and C. The $t_{\rm off}$ values enclosed by the 95\% credible regions,
however, show few differences between the two sets of models.

\begin{figure}
\includegraphics[width=\columnwidth]{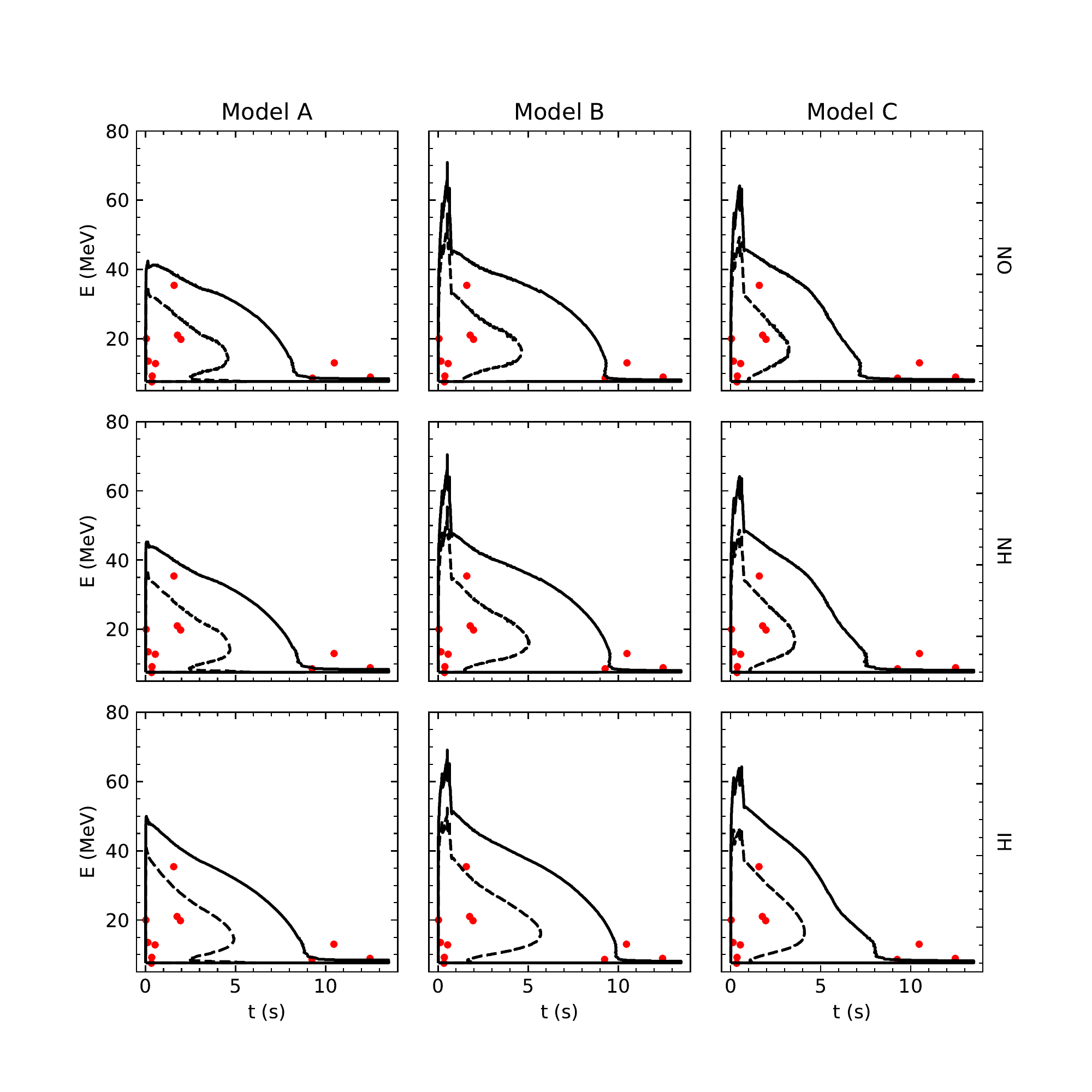}
\caption{The contours expected to enclose 68\% (dashed curve) and 95\% (solid curve) of the
events for the best-fit models A, B, and C including three cases of neutrino oscillations
are compared with the SN 1987A data (filled circles).}
\label{fig:com}
\end{figure}

Tables~\ref{tab:result} and \ref{tab:resultp} show that there is hardly any noticeable preference 
among the three cases of neutrino oscillations for each baseline model (A, B, or C) or its counterpart 
(A$'$, B$'$, or C$'$). With a Bayes factor $B_{ij}\approx34$--65, model A is strongly preferred to 
model B regardless of neutrino oscillations. The evidence in favor of model A over model C, however, 
is hardly noticeable to positive with $B_{ij}\approx2.2$--5.4. By comparison, model A$'$ is positively
preferred to model B$'$ with $B_{ij}\approx3.4$--11, while the evidence in favor of model A$'$ over
model C$'$ is hardly noticeable to positive with $B_{ij}\approx1.2$--4.4.

\section{Compatibility of Best-Fit Models with Data}
\label{sec:pvalue}
\begin{figure}
\includegraphics[width=\columnwidth]{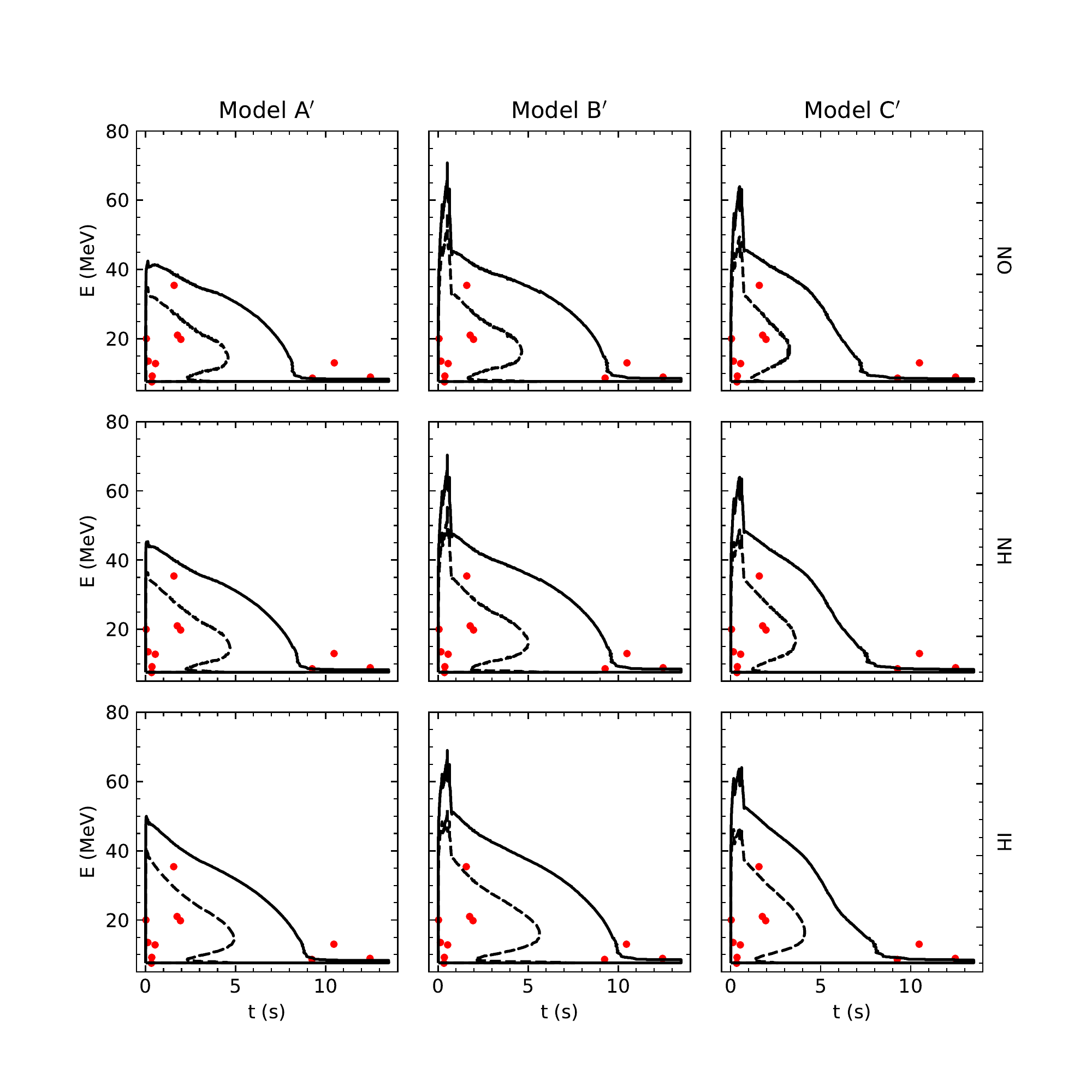}
\caption{Same as Fig.~\ref{fig:com}, but for the best-fit models A$'$, B$'$, and C$'$.}
\label{fig:comp}
\end{figure}

While the Bayesian approach can provide parameter estimates and rank the models, it does not
address the compatibility of the models with the data. To check the compatibility, we take the
frequentist approach. Specifically, we apply this approach to 
test the compatibility of the best-fit models presented
in Sec.~\ref{sec:result} with the SN 1987A data. For model $M_i$ with the set of best-fit
parameters $\boldsymbol{\theta}_{\rm bf}$, the probability of detecting an event with
energy $E$ at time $t_{\rm det}=t-t_{\rm off}$ is given by $p(E,t|\boldsymbol{\theta}_{\rm bf},M_i)$ 
[see Eq.~(\ref{eq:pet})]. Using this probability, we obtain the contours expected to enclose
68\% and 95\% of the events on the $t$-$E$ plane. These contours are shown in Figs.~\ref{fig:com}
and \ref{fig:comp} for models A, B, C and A$'$, B$'$, C$'$, respectively, along with the data.
It can be seen that for each model in the former set and its counterpart in the latter set,
the comparison of the contours with the data is very close. In addition, for all the models, 
the contours are rather consistent with the data on the eleven events.

\begin{figure}
\includegraphics[width=\columnwidth]{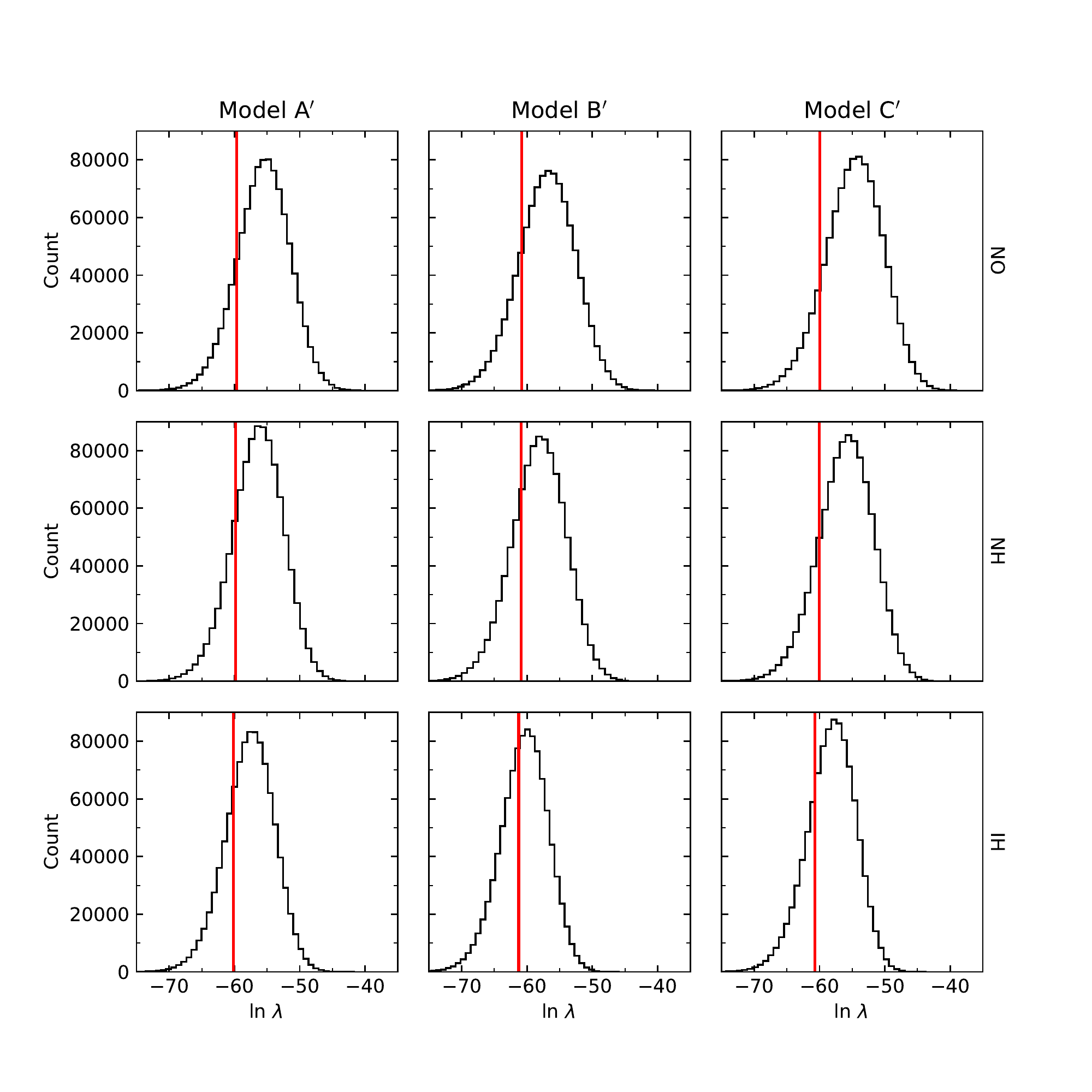}
\caption{Histograms of the statistic $\ln\lambda$ are shown for samples of
the best-fit models A$'$,  B$'$, and C$'$ including three cases of neutrino oscillations. 
A total of $10^6$ samples are simulated for each model.
The red lines denote the corresponding values of $\ln\lambda$ for the SN 1987A data.}
\label{fig:hist}
\end{figure}

For a more formal check of the compatibility, we perform a $p$-value test (see e.g., \cite{Li}), which can be
viewed as a more technical version of the comparisons shown in Figs.~\ref{fig:com} and \ref{fig:comp}.
Because the comparison of the best-fit model with the data is very close for the two sets of models,
we carry out the $p$-value tests for the best-fit models A$'$, B$'$, and C$'$ for illustration.
We draw a total of ${\cal{N}}_{\rm tot}=10^6$ samples, each consisting of $N=11$ events, from
the probability distribution $p(E,t|\boldsymbol{\theta}_{\rm bf},M_i)$ and calculate the statistic
\begin{equation}
    \lambda=\prod_{j=1}^Np(E_j,t_j|\boldsymbol{\theta}_{\rm bf},M_i)
\end{equation}
for each sample. The $p$-value is $\cal{N}_{\rm ex}/\cal{N}_{\rm tot}$, where $\cal{N}_{\rm ex}$
is the number of extreme samples with $\lambda\leq\lambda_{\rm det}$ and $\lambda_{\rm det}$ is the
statistic calculated for the detected events. Histograms of the sample test statistics 
are shown in Fig. 6, and the $p$-values for the best-fit models
A$'$, B$'$, and C$'$ are given in Table~\ref{tab:resultp}. A $p$-value smaller than 0.05 is
usually taken as evidence for rejecting the corresponding model, while a $p$-value exceeding 0.05 
simply indicates no evidence that the model is incompatible with the data. Table~\ref{tab:resultp} shows that the test yields no evidence of inconsistency between the best-fit models A$'$, B$'$, and C$'$ and the SN 1987A data, as expected from the comparisons shown in Fig.~\ref{fig:comp}.

\section{Conclusions}
\label{sec:discuss}
We have used the Bayesian approach to compare three baseline simulated models of SN neutrino emission
(see Table~\ref{tab:models} and Fig.~\ref{fig:emission}) with the KII data on SN 1987A. 
Without any modification other than inclusion of possible effects of neutrino oscillations, 
we find that model A with a brief accretion phase and a total energy of ${\cal{E}}_\nu=1.99\times 10^{53}$~ergs 
emitted in all neutrino species is the most favored. Compared to model A, model B (C) with an extended accretion 
phase and ${\cal{E}}_\nu=4.28\times 10^{53}$ ($3.30\times 10^{53}$)~ergs is strongly (barely to positively) 
disfavored. Allowing for variation of the overall neutrino flux normalization, we find that
compared to model A$'$ with a brief accretion phase, model B$'$ (C$'$) with an extended 
accretion phase is positively (barely to positively) disfavored. The best-fit model A$'$ has
${\cal{E}}_\nu\approx (2.3$--$2.6)\times 10^{53}$~ergs depending on the case of neutrino oscillations, while
the best-fit models B$'$and C$'$ both have ${\cal{E}}_\nu\approx 2.0\times 10^{53}$~ergs
regardless of neutrino oscillations (see Table~\ref{tab:resultp}). 
All the best-fit models (A, B, C, and A$'$, B$'$, C$'$) have ${\cal{E}}_\nu$ consistent with
the PNS gravitational binding energy, ${\cal{E}}_G\approx 1.34\times 10^{53}(M_G/M_\odot)^2$ erg, which is
$\sim (1.34$--$5.36)\times 10^{53}$ erg for a theoretically expected gravitational mass of
$M_G\sim 1$--$2\,M_\odot$ \cite{timmes}.
We also find no evidence that any of the best-fit models are incompatible with the data (see Figs.~\ref{fig:com} and \ref{fig:comp}
and Table~\ref{tab:resultp}).

As presented here, even with only the eleven events observed in the KII detector, 
we are able to differentiate models of neutrino emission for SN 1987A. Our analyses can be easily 
extended to models not discussed here (e.g., \cite{nagakura}). A future SN in the Galaxy is expected 
to produce $\sim 10^4$ IBD events in the Super-Kamiokande detector (e.g., \cite{scholberg}). That many events
would provide much better differentiation among models of neutrino emission for that SN. Comparing models
with data at that level would provide an important test of our understanding of the relevant physics 
and help improve SN simulations. 

\begin{acknowledgments}
We thank the Garching group for providing access to their models of SN neutrino emission.
J.O. is grateful to Ermal Rrapaj and Andre Sieverding for helpful discussions.
This work was supported in part by the US Department of Energy under grant DE-FG02-87ER40328.
Calculations were carried out at the Minnesota Supercomputing Institute.
\end{acknowledgments}


\begin{thebibliography}{25}%
\makeatletter
\providecommand \@ifxundefined [1]{%
 \@ifx{#1\undefined}
}%
\providecommand \@ifnum [1]{%
 \ifnum #1\expandafter \@firstoftwo
 \else \expandafter \@secondoftwo
 \fi
}%
\providecommand \@ifx [1]{%
 \ifx #1\expandafter \@firstoftwo
 \else \expandafter \@secondoftwo
 \fi
}%
\providecommand \natexlab [1]{#1}%
\providecommand \enquote  [1]{``#1''}%
\providecommand \bibnamefont  [1]{#1}%
\providecommand \bibfnamefont [1]{#1}%
\providecommand \citenamefont [1]{#1}%
\providecommand \href@noop [0]{\@secondoftwo}%
\providecommand \href [0]{\begingroup \@sanitize@url \@href}%
\providecommand \@href[1]{\@@startlink{#1}\@@href}%
\providecommand \@@href[1]{\endgroup#1\@@endlink}%
\providecommand \@sanitize@url [0]{\catcode `\\12\catcode `\$12\catcode
  `\&12\catcode `\#12\catcode `\^12\catcode `\_12\catcode `\%12\relax}%
\providecommand \@@startlink[1]{}%
\providecommand \@@endlink[0]{}%
\providecommand \url  [0]{\begingroup\@sanitize@url \@url }%
\providecommand \@url [1]{\endgroup\@href {#1}{\urlprefix }}%
\providecommand \urlprefix  [0]{URL }%
\providecommand \Eprint [0]{\href }%
\providecommand \doibase [0]{https://doi.org/}%
\providecommand \selectlanguage [0]{\@gobble}%
\providecommand \bibinfo  [0]{\@secondoftwo}%
\providecommand \bibfield  [0]{\@secondoftwo}%
\providecommand \translation [1]{[#1]}%
\providecommand \BibitemOpen [0]{}%
\providecommand \bibitemStop [0]{}%
\providecommand \bibitemNoStop [0]{.\EOS\space}%
\providecommand \EOS [0]{\spacefactor3000\relax}%
\providecommand \BibitemShut  [1]{\csname bibitem#1\endcsname}%
\let\auto@bib@innerbib\@empty
\bibitem [{\citenamefont {Hirata}\ \emph {et~al.}(1987)\citenamefont {Hirata}
  \emph {et~al.}}]{PhysRevLett.58.1490}%
  \BibitemOpen
  \bibfield  {author} {\bibinfo {author} {\bibfnamefont {K.}~\bibnamefont
  {Hirata}} \emph {et~al.},\ }\bibfield  {title} {\bibinfo {title} {Observation
  of a neutrino burst from the supernova sn1987a},\ }\href@noop {} {\bibfield
  {journal} {\bibinfo  {journal} {Phys. Rev. Lett.}\ }\textbf {\bibinfo
  {volume} {58}},\ \bibinfo {pages} {1490} (\bibinfo {year}
  {1987})}\BibitemShut {NoStop}%
\bibitem [{\citenamefont {Hirata}\ \emph {et~al.}(1988)\citenamefont {Hirata}
  \emph {et~al.}}]{PhysRevD.38.448}%
  \BibitemOpen
  \bibfield  {author} {\bibinfo {author} {\bibfnamefont {K.~S.}\ \bibnamefont
  {Hirata}} \emph {et~al.},\ }\bibfield  {title} {\bibinfo {title} {Observation
  in the kamiokande-ii detector of the neutrino burst from supernova sn1987a},\
  }\href@noop {} {\bibfield  {journal} {\bibinfo  {journal} {Phys. Rev. D}\
  }\textbf {\bibinfo {volume} {38}},\ \bibinfo {pages} {448} (\bibinfo {year}
  {1988})}\BibitemShut {NoStop}%
\bibitem [{\citenamefont {Bionta}\ \emph {et~al.}(1987)\citenamefont {Bionta}
  \emph {et~al.}}]{PhysRevLett.58.1494}%
  \BibitemOpen
  \bibfield  {author} {\bibinfo {author} {\bibfnamefont {R.~M.}\ \bibnamefont
  {Bionta}} \emph {et~al.},\ }\bibfield  {title} {\bibinfo {title} {Observation
  of a neutrino burst in coincidence with supernova 1987a in the large
  magellanic cloud},\ }\href@noop {} {\bibfield  {journal} {\bibinfo  {journal}
  {Phys. Rev. Lett.}\ }\textbf {\bibinfo {volume} {58}},\ \bibinfo {pages}
  {1494} (\bibinfo {year} {1987})}\BibitemShut {NoStop}%
\bibitem [{\citenamefont {Alexeyev}\ \emph {et~al.}(1988)\citenamefont
  {Alexeyev} \emph {et~al.}}]{ALEXEYEV1988209}%
  \BibitemOpen
  \bibfield  {author} {\bibinfo {author} {\bibfnamefont {E.}~\bibnamefont
  {Alexeyev}} \emph {et~al.},\ }\bibfield  {title} {\bibinfo {title} {Detection
  of the neutrino signal from sn 1987a in the lmc using the inr baksan
  underground scintillation telescope},\ }\href@noop {} {\bibfield  {journal}
  {\bibinfo  {journal} {Phys. Lett. B}\ }\textbf {\bibinfo {volume} {205}},\
  \bibinfo {pages} {209 } (\bibinfo {year} {1988})}\BibitemShut {NoStop}%
\bibitem [{\citenamefont {Arnett}\ \emph {et~al.}(1989)\citenamefont {Arnett},
  \citenamefont {Bahcall}, \citenamefont {Kirshner},\ and\ \citenamefont
  {Woosley}}]{arnett1989}%
  \BibitemOpen
  \bibfield  {author} {\bibinfo {author} {\bibfnamefont {W.~D.}\ \bibnamefont
  {Arnett}}, \bibinfo {author} {\bibfnamefont {J.~N.}\ \bibnamefont {Bahcall}},
  \bibinfo {author} {\bibfnamefont {R.~P.}\ \bibnamefont {Kirshner}},\ and\
  \bibinfo {author} {\bibfnamefont {S.~E.}\ \bibnamefont {Woosley}},\
  }\bibfield  {title} {\bibinfo {title} {Supernova 1987a},\ }\href@noop {}
  {\bibfield  {journal} {\bibinfo  {journal} {Annu. Rev. Astron. Astrophys.}\
  }\textbf {\bibinfo {volume} {27}},\ \bibinfo {pages} {629} (\bibinfo {year}
  {1989})}\BibitemShut {NoStop}%
\bibitem [{\citenamefont {Loredo}\ and\ \citenamefont
  {Lamb}(2002)}]{Loredo:2001rx}%
  \BibitemOpen
  \bibfield  {author} {\bibinfo {author} {\bibfnamefont {T.~J.}\ \bibnamefont
  {Loredo}}\ and\ \bibinfo {author} {\bibfnamefont {D.~Q.}\ \bibnamefont
  {Lamb}},\ }\bibfield  {title} {\bibinfo {title} {{Bayesian analysis of
  neutrinos observed from supernova SN 1987A}},\ }\href@noop {} {\bibfield
  {journal} {\bibinfo  {journal} {Phys. Rev. D}\ }\textbf {\bibinfo {volume}
  {65}},\ \bibinfo {pages} {063002} (\bibinfo {year} {2002})}\BibitemShut
  {NoStop}%
\bibitem [{\citenamefont {Costantini}\ \emph {et~al.}(2007)\citenamefont
  {Costantini}, \citenamefont {Ianni}, \citenamefont {Pagliaroli},\ and\
  \citenamefont {Vissani}}]{Costantini_2007}%
  \BibitemOpen
  \bibfield  {author} {\bibinfo {author} {\bibfnamefont {M.~L.}\ \bibnamefont
  {Costantini}}, \bibinfo {author} {\bibfnamefont {A.}~\bibnamefont {Ianni}},
  \bibinfo {author} {\bibfnamefont {G.}~\bibnamefont {Pagliaroli}},\ and\
  \bibinfo {author} {\bibfnamefont {F.}~\bibnamefont {Vissani}},\ }\bibfield
  {title} {\bibinfo {title} {Is there a problem with low energy {SN}1987a
  neutrinos?},\ }\href@noop {} {\bibfield  {journal} {\bibinfo  {journal} {J.
  Cosmo. Astropart. Phys.}\ }\textbf {\bibinfo {volume} {2007}},\ \bibinfo
  {pages} {014} (\bibinfo {year} {2007})}\BibitemShut {NoStop}%
\bibitem [{\citenamefont {{Janka}}(2012)}]{janka2012}%
  \BibitemOpen
  \bibfield  {author} {\bibinfo {author} {\bibfnamefont {H.-T.}\ \bibnamefont
  {{Janka}}},\ }\bibfield  {title} {\bibinfo {title} {{Explosion mechanisms of
  core-collapse supernovae}},\ }\href@noop {} {\bibfield  {journal} {\bibinfo
  {journal} {Annu. Rev. Nucl. Part. Sci.}\ }\textbf {\bibinfo {volume} {62}},\
  \bibinfo {pages} {407} (\bibinfo {year} {2012})}\BibitemShut {NoStop}%
\bibitem [{\citenamefont {Mirizzi}\ \emph {et~al.}(2016)\citenamefont
  {Mirizzi}, \citenamefont {Tamborra}, \citenamefont {Janka}, \citenamefont
  {Saviano}, \citenamefont {Scholberg}, \citenamefont {Bollig}, \citenamefont
  {Hudepohl},\ and\ \citenamefont {Chakraborty}}]{Mirizzi:2015eza}%
  \BibitemOpen
  \bibfield  {author} {\bibinfo {author} {\bibfnamefont {A.}~\bibnamefont
  {Mirizzi}}, \bibinfo {author} {\bibfnamefont {I.}~\bibnamefont {Tamborra}},
  \bibinfo {author} {\bibfnamefont {H.-T.}\ \bibnamefont {Janka}}, \bibinfo
  {author} {\bibfnamefont {N.}~\bibnamefont {Saviano}}, \bibinfo {author}
  {\bibfnamefont {K.}~\bibnamefont {Scholberg}}, \bibinfo {author}
  {\bibfnamefont {R.}~\bibnamefont {Bollig}}, \bibinfo {author} {\bibfnamefont
  {L.}~\bibnamefont {Hudepohl}},\ and\ \bibinfo {author} {\bibfnamefont
  {S.}~\bibnamefont {Chakraborty}},\ }\bibfield  {title} {\bibinfo {title}
  {{Supernova neutrinos: Production, oscillations and detection}},\ }\href@noop
  {} {\bibfield  {journal} {\bibinfo  {journal} {Riv. Nuovo Cim.}\ }\textbf
  {\bibinfo {volume} {39}},\ \bibinfo {pages} {1} (\bibinfo {year}
  {2016})}\BibitemShut {NoStop}%
\bibitem [{Gar()}]{Garching}%
  \BibitemOpen
  \href@noop {} {}\bibinfo {howpublished}
  {https://wwwmpa.mpa-garching.mpg.de/ccsnarchive/}\BibitemShut {NoStop}%
\bibitem [{\citenamefont {{Steiner}}\ \emph {et~al.}(2013)\citenamefont
  {{Steiner}}, \citenamefont {{Hempel}},\ and\ \citenamefont
  {{Fischer}}}]{SFHo}%
  \BibitemOpen
  \bibfield  {author} {\bibinfo {author} {\bibfnamefont {A.~W.}\ \bibnamefont
  {{Steiner}}}, \bibinfo {author} {\bibfnamefont {M.}~\bibnamefont
  {{Hempel}}},\ and\ \bibinfo {author} {\bibfnamefont {T.}~\bibnamefont
  {{Fischer}}},\ }\bibfield  {title} {\bibinfo {title} {{Core-collapse
  supernova equations of state based on neutron star observations}},\
  }\href@noop {} {\bibfield  {journal} {\bibinfo  {journal} {Astrophys. J.}\
  }\textbf {\bibinfo {volume} {774}},\ \bibinfo {eid} {17} (\bibinfo {year}
  {2013})}\BibitemShut {NoStop}%
\bibitem [{\citenamefont {{Lattimer}}\ and\ \citenamefont
  {{Swesty}}(1991)}]{LS220}%
  \BibitemOpen
  \bibfield  {author} {\bibinfo {author} {\bibfnamefont {J.~M.}\ \bibnamefont
  {{Lattimer}}}\ and\ \bibinfo {author} {\bibfnamefont {D.~F.}\ \bibnamefont
  {{Swesty}}},\ }\bibfield  {title} {\bibinfo {title} {{A generalized equation
  of state for hot, dense matter}},\ }\href@noop {} {\bibfield  {journal}
  {\bibinfo  {journal} {Nucl. Phys. A}\ }\textbf {\bibinfo {volume} {535}},\
  \bibinfo {pages} {331} (\bibinfo {year} {1991})}\BibitemShut {NoStop}%
\bibitem [{\citenamefont {Tamborra}\ \emph {et~al.}(2012)\citenamefont
  {Tamborra}, \citenamefont {Mueller}, \citenamefont {Huedepohl}, \citenamefont
  {Janka},\ and\ \citenamefont {Raffelt}}]{Tamborra2012}%
  \BibitemOpen
  \bibfield  {author} {\bibinfo {author} {\bibfnamefont {I.}~\bibnamefont
  {Tamborra}}, \bibinfo {author} {\bibfnamefont {B.}~\bibnamefont {Mueller}},
  \bibinfo {author} {\bibfnamefont {L.}~\bibnamefont {Huedepohl}}, \bibinfo
  {author} {\bibfnamefont {H.-T.}\ \bibnamefont {Janka}},\ and\ \bibinfo
  {author} {\bibfnamefont {G.}~\bibnamefont {Raffelt}},\ }\bibfield  {title}
  {\bibinfo {title} {High-resolution supernova neutrino spectra represented by
  a simple fit},\ }\href@noop {} {\bibfield  {journal} {\bibinfo  {journal}
  {Phys. Rev. D}\ }\textbf {\bibinfo {volume} {86}},\ \bibinfo {pages} {125031}
  (\bibinfo {year} {2012})}\BibitemShut {NoStop}%
\bibitem [{\citenamefont {{Panagia}}\ \emph {et~al.}(1991)\citenamefont
  {{Panagia}}, \citenamefont {{Gilmozzi}}, \citenamefont {{Macchetto}},
  \citenamefont {{Adorf}},\ and\ \citenamefont
  {{Kirshner}}}]{1991ApJ...380L..23P}%
  \BibitemOpen
  \bibfield  {author} {\bibinfo {author} {\bibfnamefont {N.}~\bibnamefont
  {{Panagia}}}, \bibinfo {author} {\bibfnamefont {R.}~\bibnamefont
  {{Gilmozzi}}}, \bibinfo {author} {\bibfnamefont {F.}~\bibnamefont
  {{Macchetto}}}, \bibinfo {author} {\bibfnamefont {H.~M.}\ \bibnamefont
  {{Adorf}}},\ and\ \bibinfo {author} {\bibfnamefont {R.~P.}\ \bibnamefont
  {{Kirshner}}},\ }\bibfield  {title} {\bibinfo {title} {{Properties of the SN
  1987A circumstellar ring and the distance to the large magellanic cloud}},\
  }\href@noop {} {\bibfield  {journal} {\bibinfo  {journal} {Astrophys. J.}\
  }\textbf {\bibinfo {volume} {380}},\ \bibinfo {pages} {L23} (\bibinfo {year}
  {1991})}\BibitemShut {NoStop}%
\bibitem [{\citenamefont {Panagia}(2005)}]{Panagia:2003rt}%
  \BibitemOpen
  \bibfield  {author} {\bibinfo {author} {\bibfnamefont {N.}~\bibnamefont
  {Panagia}},\ }\bibfield  {title} {\bibinfo {title} {{A geometric
  determination of the distance to SN 1987A and the LMC}},\ }\href@noop {}
  {\bibfield  {journal} {\bibinfo  {journal} {Springer Proc. Phys.}\ }\textbf
  {\bibinfo {volume} {99}},\ \bibinfo {pages} {585} (\bibinfo {year}
  {2005})}\BibitemShut {NoStop}%
\bibitem [{\citenamefont {Dighe}\ and\ \citenamefont
  {Smirnov}(2000)}]{PhysRevD.62.033007}%
  \BibitemOpen
  \bibfield  {author} {\bibinfo {author} {\bibfnamefont {A.~S.}\ \bibnamefont
  {Dighe}}\ and\ \bibinfo {author} {\bibfnamefont {A.~Y.}\ \bibnamefont
  {Smirnov}},\ }\bibfield  {title} {\bibinfo {title} {Identifying the neutrino
  mass spectrum from a supernova neutrino burst},\ }\href
  {https://doi.org/10.1103/PhysRevD.62.033007} {\bibfield  {journal} {\bibinfo
  {journal} {Phys. Rev. D}\ }\textbf {\bibinfo {volume} {62}},\ \bibinfo
  {pages} {033007} (\bibinfo {year} {2000})}\BibitemShut {NoStop}%
\bibitem [{\citenamefont {{Gonzalez-Garcia}}\ \emph {et~al.}(2014)\citenamefont
  {{Gonzalez-Garcia}}, \citenamefont {{Maltoni}},\ and\ \citenamefont
  {{Schwetz}}}]{Gonzalez-Garcia}%
  \BibitemOpen
  \bibfield  {author} {\bibinfo {author} {\bibfnamefont {M.~C.}\ \bibnamefont
  {{Gonzalez-Garcia}}}, \bibinfo {author} {\bibfnamefont {M.}~\bibnamefont
  {{Maltoni}}},\ and\ \bibinfo {author} {\bibfnamefont {T.}~\bibnamefont
  {{Schwetz}}},\ }\bibfield  {title} {\bibinfo {title} {{Updated fit to three
  neutrino mixing: Status of leptonic CP violation}},\ }\href@noop {}
  {\bibfield  {journal} {\bibinfo  {journal} {J. High Energy Phys.}\ }\textbf
  {\bibinfo {volume} {2014}},\ \bibinfo {eid} {52}}\BibitemShut {NoStop}%
\bibitem [{\citenamefont {Zyla}\ \emph {et~al.}(2020)\citenamefont {Zyla} \emph
  {et~al.}}]{Zyla:2020zbs}%
  \BibitemOpen
  \bibfield  {author} {\bibinfo {author} {\bibfnamefont {P.}~\bibnamefont
  {Zyla}} \emph {et~al.} (\bibinfo {collaboration} {Particle Data Group}),\
  }\bibfield  {title} {\bibinfo {title} {{Review of Particle Physics}},\ }\href
  {https://doi.org/10.1093/ptep/ptaa104} {\bibfield  {journal} {\bibinfo
  {journal} {PTEP}\ }\textbf {\bibinfo {volume} {2020}},\ \bibinfo {pages}
  {083C01} (\bibinfo {year} {2020})}\BibitemShut {NoStop}%
\bibitem [{\citenamefont {Jegerlehner}\ \emph {et~al.}(1996)\citenamefont
  {Jegerlehner}, \citenamefont {Neubig},\ and\ \citenamefont
  {Raffelt}}]{PhysRevD.54.1194}%
  \BibitemOpen
  \bibfield  {author} {\bibinfo {author} {\bibfnamefont {B.}~\bibnamefont
  {Jegerlehner}}, \bibinfo {author} {\bibfnamefont {F.}~\bibnamefont
  {Neubig}},\ and\ \bibinfo {author} {\bibfnamefont {G.}~\bibnamefont
  {Raffelt}},\ }\bibfield  {title} {\bibinfo {title} {Neutrino oscillations and
  the supernova 1987a signal},\ }\href@noop {} {\bibfield  {journal} {\bibinfo
  {journal} {Phys. Rev. D}\ }\textbf {\bibinfo {volume} {54}},\ \bibinfo
  {pages} {1194} (\bibinfo {year} {1996})}\BibitemShut {NoStop}%
\bibitem [{\citenamefont {Lunardini}\ and\ \citenamefont
  {Smirnov}(2004)}]{55af5131dc954e7592bb1d948c16becd}%
  \BibitemOpen
  \bibfield  {author} {\bibinfo {author} {\bibfnamefont {C.}~\bibnamefont
  {Lunardini}}\ and\ \bibinfo {author} {\bibfnamefont {A.}~\bibnamefont
  {Smirnov}},\ }\bibfield  {title} {\bibinfo {title} {Neutrinos from sn1987a:
  Flavor conversion and interpretation of results},\ }\href@noop {} {\bibfield
  {journal} {\bibinfo  {journal} {Astropart. Phys.}\ }\textbf {\bibinfo
  {volume} {21}},\ \bibinfo {pages} {703} (\bibinfo {year} {2004})}\BibitemShut
  {NoStop}%
\bibitem [{\citenamefont {Barlow}(1990)}]{BARLOW1990496}%
  \BibitemOpen
  \bibfield  {author} {\bibinfo {author} {\bibfnamefont {R.}~\bibnamefont
  {Barlow}},\ }\bibfield  {title} {\bibinfo {title} {Extended maximum
  likelihood},\ }\href@noop {} {\bibfield  {journal} {\bibinfo  {journal}
  {Nucl. Instrum. Methods Phys. Res. A}\ }\textbf {\bibinfo {volume} {297}},\
  \bibinfo {pages} {496 } (\bibinfo {year} {1990})}\BibitemShut {NoStop}%
\bibitem [{\citenamefont {Li}(2017)}]{Li}%
  \BibitemOpen
  \bibfield  {author} {\bibinfo {author} {\bibfnamefont {C.-H.}\ \bibnamefont
  {Li}},\ }\emph {\bibinfo {title} {Astrophysics and physics of neutrino
  detection}},\ \href@noop {} {Ph.D. thesis},\ \bibinfo  {school} {University
  of Minnesota} (\bibinfo {year} {2017}),\ \bibinfo {note} {retrieved from the
  University of Minnesota Digital Conservancy,
  http://hdl.handle.net/11299/191310.}\BibitemShut {Stop}%
\bibitem [{\citenamefont {{Timmes}}\ \emph {et~al.}(1996)\citenamefont
  {{Timmes}}, \citenamefont {{Woosley}},\ and\ \citenamefont
  {{Weaver}}}]{timmes}%
  \BibitemOpen
  \bibfield  {author} {\bibinfo {author} {\bibfnamefont {F.~X.}\ \bibnamefont
  {{Timmes}}}, \bibinfo {author} {\bibfnamefont {S.~E.}\ \bibnamefont
  {{Woosley}}},\ and\ \bibinfo {author} {\bibfnamefont {T.~A.}\ \bibnamefont
  {{Weaver}}},\ }\bibfield  {title} {\bibinfo {title} {{The Neutron Star and
  Black Hole Initial Mass Function}},\ }\href@noop {} {\bibfield  {journal}
  {\bibinfo  {journal} {Astrophys. J.}\ }\textbf {\bibinfo {volume} {457}},\
  \bibinfo {pages} {834} (\bibinfo {year} {1996})}\BibitemShut {NoStop}%
\bibitem [{\citenamefont {{Nagakura}}\ \emph {et~al.}(2021)\citenamefont
  {{Nagakura}}, \citenamefont {{Burrows}},\ and\ \citenamefont
  {{Vartanyan}}}]{nagakura}%
  \BibitemOpen
  \bibfield  {author} {\bibinfo {author} {\bibfnamefont {H.}~\bibnamefont
  {{Nagakura}}}, \bibinfo {author} {\bibfnamefont {A.}~\bibnamefont
  {{Burrows}}},\ and\ \bibinfo {author} {\bibfnamefont {D.}~\bibnamefont
  {{Vartanyan}}},\ }\bibfield  {title} {\bibinfo {title} {{Supernova neutrino
  signals based on long-term axisymmetric simulations}},\ }\href@noop {}
  {\bibfield  {journal} {\bibinfo  {journal} {Mon. Not. Roy. Astron. Soc.}\
  }\textbf {\bibinfo {volume} {506}},\ \bibinfo {pages} {1462} (\bibinfo {year}
  {2021})}\BibitemShut {NoStop}%
\bibitem [{\citenamefont {Scholberg}(2012)}]{scholberg}%
  \BibitemOpen
  \bibfield  {author} {\bibinfo {author} {\bibfnamefont {K.}~\bibnamefont
  {Scholberg}},\ }\bibfield  {title} {\bibinfo {title} {Supernova neutrino
  detection},\ }\href@noop {} {\bibfield  {journal} {\bibinfo  {journal} {Annu.
  Rev. Nucl. Part. Sci.}\ }\textbf {\bibinfo {volume} {62}},\ \bibinfo {pages}
  {81} (\bibinfo {year} {2012})}\BibitemShut {NoStop}%
\end{thebibliography}

%

\end{document}